\documentclass[acmlarge]{acmart}

\makeatletter
\newcommand{\myconfshort}{\acmConference@shortname}
\newcommand{\myconffull}{\acmConference@name}
\newcommand{\myconfdate}{\acmConference@date}
\newcommand{\myconfloc}{\acmConference@venue}
\AtBeginDocument{
  \fancypagestyle{firstpagestyle}{
    \fancyhead{}%
    \fancyfoot[C]{}%
  }
  \fancyhf{}
  \fancyhead[LO]{\@headfootfont\shorttitle}%
  \fancyhead[RE]{\@headfootfont\@shortauthors}%
  \fancyhead[LE]{\@headfootfont\footnotesize \myconfshort, \myconfdate, \myconfloc}%
  \fancyhead[RO]{\@headfootfont\footnotesize \myconfshort, \myconfdate, \myconfloc}%
  \fancyfoot[C]{}%
}
\makeatother
\acmBooktitle{\conffull\@ (\confshort), \confdate, \confloc}

\AtBeginDocument{%
  }

\usepackage{hyperref}

\usepackage{tcolorbox}
\setcopyright{acmlicensed}

\copyrightyear{2026}
\acmYear{2026}
\setcopyright{cc}
\setcctype{by}
\acmConference[FAccT '26]{The 2026 ACM Conference on Fairness, Accountability, and Transparency}{June 25--28, 2026}{Montreal, QC, Canada}
\acmBooktitle{The 2026 ACM Conference on Fairness, Accountability, and Transparency (FAccT '26), June 25--28, 2026, Montreal, QC, Canada}
\acmDOI{10.1145/3805689.3812227}
\acmISBN{979-8-4007-2596-8/2026/06}

\begin{document}

% \title{Mapping Data Labour Supply Chain in Africa in an Era of Digital Apartheid: a Struggle for Recognition}
\title[Mapping Data Labour Supply Chain in Africa in an Era of Digital Apartheid]{Mapping Data Labour Supply Chain in Africa in an Era of Digital Apartheid: a Struggle for Recognition}

\author{Jessica Pidoux}
\authornote{Equal contribution}
\email{jessica.pidoux@unine.ch}
\orcid{0000-0001-5705-6230}
\affiliation{%
  \institution{University of Neuchâtel}
  \country{Switzerland}}
\affiliation{%
  \institution{PersonalData.IO}
  \country{Switzerland}}
\author{Mariame Tighanimine}
\email{mariame@tighanimine.com}
\orcid{0009-0005-8188-9976}
\authornotemark[1]
 \affiliation{%
  \institution{University of Neuchâtel}
  \country{Switzerland}}
\affiliation{%
  \institution{PersonalData.IO}
  \country{Switzerland}}
\affiliation{%
  \institution{Lise Cnam CNRS}
  \country{France}}
% \authornotemark[2] 
%\authornote{Also affiliated with Lise, Cnam CNRS, France.}
\authornote{Corresponding author: mariame@tighanimine.com}
% \affiliation{%
%   \institution{Lise, Cnam CNRS}
%   \country{France}}

% \author{Jauthor1}
% \email{j}
% % \orcid{1234-5678-9012}
% \author{author2}
% \email{k}
% \affiliation{%
%   \institution{University of Neuchâtel}
%   \country{Switzerland}}
% \affiliation{%
%   \institution{PersonalData.IO}
%   \country{Switzerland}}
% \authornotemark[2]  % Additional note for the second affiliation
% \affiliation{%
%   \institution{Another Institution Name}
%   \country{Another Country}}

% \author{Jessica Pidoux}
% \email{jessica.pidoux@unine.ch}
% \affiliation{%
%   \institution{University of Neuchâtel}
%   \country{Switzerland}}
% \affiliation{%
%   \institution{PersonalData.IO}
%   \country{Switzerland}}

% \author{Mariame Tighanimine}
% \email{mariame@tighanimine.com}
% \affiliation{%
%   \institution{University of Neuchâtel}
%   \country{Switzerland}}
% \affiliation{%
%   \institution{PersonalData.IO}
%   \country{Switzerland}}
% \affiliation{%
%   \institution{Lise, Cnam CNRS}
%   \country{France}}

\author{Sofia Kypraiou}
\email{sofia.kypraiou@personaldata.io}
\orcid{0000-0003-4656-8446}
\affiliation{%
  \institution{University of Neuchâtel}
  \country{Switzerland}}
\affiliation{%
  \institution{PersonalData.IO}
  \country{Switzerland}}

\author{Sonia Kgomo}
\email{soniamatete35@gmail.com}
\orcid{0009-0008-7865-6962}
\affiliation{%
  \institution{African Content Moderators Union}
  \country{Kenya}}

\author{Kauna Ibrahim Malgwi}
\email{kaunamalgwi@gmail.com}
\orcid{0009-0008-9685-740X}
\affiliation{%
  \institution{African Content Moderators Union}
  \country{Kenya}}
\affiliation{%
  \institution{Kenya Digital Rights and Mental Health Initiative}
  \country{Kenya}}

\author{Richard Mwaura Mathenge}
\email{richiemathenge9@gmail.com}
\orcid{0009-0004-7100-0259}
\affiliation{%
  \institution{African Content Moderators Union}
  \country{Kenya}}
\affiliation{%
  \institution{Kenya Techworker Community Africa}
  \country{Kenya}}

\author{Mophat Okinyi}
\email{okinyi.mophat@gmail.com}
\orcid{0009-0000-5932-3232}
\affiliation{%
  \institution{African Content Moderators Union}
  \country{Kenya}}
\affiliation{%
  \institution{Kenya Techworker Community Africa}
  \country{Kenya}}

\author{James Oyange}
\email{oyangej8@gmail.com}
\orcid{0009-0009-4917-710X}
\affiliation{%
  \institution{African Content Moderators Union}
  \country{Kenya}}

\renewcommand{\shortauthors}{Pidoux et al.}

\begin{abstract}

Content moderation and data annotation work has shifted to the Global South, particularly Africa, where workers at business process outsourcing (BPO) companies operate under precarity to serve Global North needs. We address the invisibility of this data labour supply chain and the underdocumented working conditions of its workforce. Drawing on a participatory collaboration between academics, an NGO, and a union, we conducted desk research and deployed a questionnaire (n=81) attuned to unions' organising goals. Our findings show that data labour spans 43 out of 55 African countries, involving 17 major firms serving predominantly North-American and European clients, with workers employed on short-term contracts, under psychological stress and economic instability — conditions that obscure the competences, i.e. adaptability and resilience, that their work demands. We contribute the first comprehensive map of Africa’s data labour industry and demonstrate a methodology that centers workers’ collective actions in documenting their conditions, drawing on Honneth’s ``struggle for recognition'' to capture workers’ demands for professional and social acknowledgement.

\end{abstract}





\maketitle

\section{Introduction}

The proliferation of artificial intelligence (AI) applications which are developed predominantly by North American technology companies, have generated new forms of human labour collectively termed ``data work'' that is rendered invisible \cite{miceli2022data, lampinen2024work}. This work encompasses tasks like filtering hate speech, classifying videos, images, or tagging text through data labelling, dataset preparation, and so, performed by workers employed by intermediary outsourcing companies, whose contractual arrangements are often precarious \cite{tubaro2025does}. Sama, a San Francisco-based business process outsourcing (henceforth BPO) company operating across the Global South, particularly in Africa, exemplifies these emerging labour arrangements. Despite describing itself as an ``ethical AI company'', Sama has faced significant scrutiny following investigative reporting. A Time investigation \cite{perrigo2023exclusive} showed that Kenyan employees earned less than two dollars per hour for reviewing and classifying disturbing content to train ChatGPT's systems without any professional safeguard. Fairwork documented ``a broader culture of fear'' within Sama \cite{cant2023fairwork}. While Fairwork acknowledges that Sama's management engaged with their findings and expressed commitment to improving working conditions, ongoing legal challenges and continued worker mobilisation indicate that these systemic problems persist \cite{agbetiloye}. 
\\
To our knowledge, there is no landscape of the organisations that constitute the outsourcing data work in Africa with respect to workers' accounting of their working conditions. We address this gap by offering a double perspective on the AI industry and its underlying working conditions as experienced by workers. This double macro-micro level perspective allows to investigate how the AI industry is transforming divisions of labour (where AI is developed, and trained), work value, and skills recognition, through participatory research in collaboration with an NGO and a union of African data workers.
\\
These workers have been actively denouncing their working conditions following their dismissal and pursuing legal and advocacy actions for professional recognition. Importantly, data workers do not reject the nature of their work itself, but rather contest the exploitative conditions under which they labour. Their approach centres on gaining recognition for their expertise, qualifications, competencies, and status as essential contributors to the AI pipeline in order to improve their working conditions. Their actions are part of what sociologist Axel Honneth terms a ``struggle for recognition.'' \\
This struggle for recognition occurs within a broader political context concerning how labour is divided globally and how workers can participate democratically in shaping their working conditions. However, the current context of information system design in which data workers are a key workforce is affected by a ``black tide'' that primarily harms the populations of the Horn of Africa \cite{gebru2023digitalapartheid}, who are directly affected by the harmful effects of the tech industry of the Global North. These effects include both the consequences of online mis/disinformation (in the form of calls for genocide, human trafficking, and the spread of unmoderated or poorly moderated hate speech) and the extraction of natural and human resources to guarantee western populations a better experience of the Internet and its potential. African workers, to use the metaphor of researchers Timnit Gebru et al.~\cite{gebru2023digitalapartheid} ``clean up the oil spill (...) at the expense of their own well being.'' This is what AI security researcher El-Mahdi El-Mhamdi calls ``digital apartheid'' \cite{gebru2023digitalapartheid, elmahdielmhamdi2021}. 
\\
In such digital apartheid, mapping work and data flows in relation to their working conditions in Africa helps to illustrate how segregative dynamics that are expressed in different worlds of work \cite{chappe2024dynamiques} are part of the macroeconomic developments of contemporary capitalism beyond the attention that research has given to the issues of data work. Indeed, these macrodynamics draw social, ethnic, and moral boundaries \cite{lamont2002study} between different categories of humans in general and workers in particular. At the same time, taking a close look at the willingness of a group of workers involved in a participatory study, fighting for recognition, allows us to identify the effects of these segregative dynamics on a microsociological scale. It also allows us to move beyond dominant media and academic discourse that characterises these workers as ``invisible'', relegated to ``dirty work'' \cite{hughes1951work}, or labels them as ``underclass'' or ``lumpenproletariat.'' While such characterisations may capture certain aspects of their situation, this discourse can inadvertently undermine data workers' struggle for recognition of their working conditions.
This paper makes three contributions. First, it produces the first visual materialisation of the AI data work supply chain in Africa, mapping 17 firms operating across 43 of Africa's 55 countries and serving predominantly North American and European clients, making visible a chain of labour relations that has remained opaque due to industry secrecy. Second, it provides systematic empirical evidence of data workers' conditions through a questionnaire (n=81), documenting patterns of precarity — short-term contracts, wages below national averages, routine overwork, and structural silencing — that are consistent across companies and countries. Third, it demonstrates a participatory methodology that positions workers as active co-producers of scientific knowledge, combining their insider expertise with academic analysis to document working conditions for union advocacy, and advancing a broader argument for the recognition of data work as a skilled occupation with corresponding labour protections.
\section{Context: The Content Moderators Union Struggle for Recognition}

The African Content Moderators Union is a collective of workers denouncing abusive working conditions \cite{musambi2023facebook} such as job insecurity, surveillance, arbitrary dismissal, and wellbeing harms in the content moderation industry. They united with hundreds of other content moderators in Africa and globally, pursuing legal cases against Meta for exploitative conditions via their subcontractors’ companies like Sama \cite{restofworldMetaModerators, siliconsavanna, guardianFacebookModerators, independentKenyanModerators}. The union is now recognised for speaking out publicly and claiming social justice after being isolated and censured by confidentiality clauses at their subcontracting companies. Hence, this union and their struggle represent a major social movement of emerging workers in the AI sector that are crucial for technology development but have not  been recognised as such.

Honneth's ``struggle for recognition'' concept \cite{honneth1996struggle} allows to situate data workers' mobilisation within a tradition of labour movements challenging conditions of invisibility and exploitation. It is relevant for understanding how data workers in Africa navigate undemocratic or illiberal national contexts while fighting for recognition of their qualifications. Rather than viewing these workers solely through the lens of exploitation, Honneth's framework allows us to examine how workers assert their professional identities and competencies as forms of resistance, encompassing both acknowledgment of their technical skills and demands for dignified working conditions that would enable them to fully exercise their expertise and advance their careers.

Since the end of the 20th century, recognition has been expressed in numerous social struggles and has been the subject of interest in social sciences, particularly those related to labour. From the very beginning, scholarship on recognition has postulated that the transition to a post-Fordist and post-socialist world has led to a shift beyond simple criticism of the economic structure of society to consider social injustices, particularly cultural ones. Recognition has become an important concept for understanding the social and political transformations of the 20th century \cite{butler2002gender, taylor2021politics, fraser2020redistribution, kymlicka1995multicultural}. Honneth goes further in his conception of recognition \cite{honneth1992integrity} and focuses specifically on work and its importance in the social world as we experience it \cite{honneth2013wir}. In other words, the social esteem received by the participants in a work process determines the possibility of self-esteem and awareness of one's own value. This shift in focus, from identity to the recognition of work, qualities, and ultimately workers themselves, raises questions about working conditions, which are overall improved material conditions of existence. In the case of data workers, the aim is therefore to understand the conditions of production within the AI supply chain.
\section{Related Work}

Data work is a key component of an international organisation and division of labour within the AI sector. Muldoon et al. \cite{muldoon2025politics} situate it in the global AI supply chains, stemming in particular from AI preparation, which in turn flows from AI infrastructure. AI data work done within BPOs, despite its significance in the ever-accelerating development of AI systems, still receives only limited attention from researchers. Our work is intended to form part of this emerging literature.

\subsection{Business Process Outsourcing Companies and Global South Data Labour Arrangements}

When examining data labour arrangements, crowdsourcing platforms are more widely recognised than BPO arrangements. Early forms of ``microwork'', ``crowdwork'', or ``data work'' \cite{muldoon2024typology} initially flourished through digital platforms that facilitated online matching between supply and demand for small temporary jobs \cite{ross2010crowdworkers, denton2021genealogy, tubaro2025does}. However, the labour supply chains used for AI development have evolved in recent years towards more complex and diverse forms of organisation and workforce provision \cite{miceli2020between, miceli2022data, schmidt2022planetary, lindquist2022follower, muldoon2025poverty}. BPOs represent one such organisational model. 

BPOs are long-established organisations that provide a wide range of services, including human resources, finance, accounting, and after-sales support. BPOs can be understood as a subset of outsourcing arrangements, with call centres representing one of the most prominent examples of this model \cite{apte1995global, miller2018call}. An important distinction between BPOs and crowdsourcing platforms lies in their management structures: BPOs involve human management \cite{le2023problem}, while crowdsourcing platforms rely on algorithmic management of workers. Within the AI industry, the BPO model is regarded as an effective approach to integrate workers into production processes \cite{miceli2020between}. This integration is particularly valued given the dynamic nature of data annotation work, especially content moderation, and BPOs' established experience in delivering diverse services to tertiary sector companies. However, this BPO model is detrimental to the professionalisation of workers. Content moderation work generates value for platforms beyond its primary function of managing user-generated content. The decisions made by content moderators create structured datasets that are used to train automated detection systems and machine learning models \cite{gebrekidan2024content, abdelkadir2025role}. These AI systems become integral to platforms' ability to scale content management and form the foundation for algorithmic recommendation systems that drive user engagement and advertising revenue \cite{gillespie2018custodians}. Content moderators, and more broadly data workers, working within BPO arrangements often receive limited recognition for these contributions within formal organisational structures, despite their work being essential to multiple revenue-generating activities. BPOs geographical locations are often selected based on labour costs \cite{borman2006applying} and historical ties between contracting companies and host countries, frequently reflecting colonial relationships. France and Madagascar exemplify this pattern \cite{le2023problem}. 

Studies examining how these organisations affect minority groups disadvantaged by platform moderation mechanisms \cite{haimson2021disproportionate} or countries in the Global South \cite{shahid2023decolonizing} remain limited. Global South populations are disproportionately affected by disinformation, hate speech, and incitement to violence, including cases of genocide incitement in Myanmar and Tigray \cite{fink2018dangerous, brooten2020media, nkemelu2022tackling, wilmot2021dueling, crystal2023facebook}, whilst platforms allocate minimal or no moderation resources to these regions \cite{abdelkadir2025role}, speaking low-resourced languages \cite{ghosh2023chatgpt, nigatu2024searched}. This is particularly paradoxical given that BPOs established in these same countries perform content moderation on behalf of Global North platforms, primarily serving Anglo-Saxon markets. This asymmetry not only exposes local populations to unmoderated harm but further burdens the working conditions of moderators themselves, who must navigate guidelines designed for Anglo-Saxon contexts that fail to account for their own social, political, and linguistic realities, leaving them without adequate frameworks to handle the content they encounter \cite{abdelkadir2025role}.

\subsection{What Counts as Data Work? Definitions, Recognition, and Working Conditions}

Content moderation can be generally defined as ``the organized practice of screening user-generated content posted to Internet sites, social media and other online outlets, in order to determine the appropriateness of the content for a given site, locality, or jurisdiction \cite{roberts2014behind}'' (cited in \cite{moran2025end}). Other authors explain that content moderation represents an applied and specialised form of data annotation activity that occurs in near real-time \cite{abdelkadir2025role}. However, authors highlight that content moderation extends beyond traditional data labelling, typically referring to the process of assigning labels or tags to data, by requiring workers to navigate dynamic, uncertain, and complex textual, technical, and social contexts \cite{kapania2023hunt, gebrekidan2024content, jain2024filter, benwraycontent2025}. This makes moderation work inherently unpredictable, as situations frequently exceed the frameworks established by platform guidelines. 

Although existing research has begun to document the contextual, linguistic, and cultural knowledge that moderation requires, definitions of data work share a common limitation: they focus primarily on task categorisation rather than on the broader competencies workers bring to their roles and the occupational recognition these should command. This gap between how data work is defined and how it is practised by workers remains underexplored, and constitutes the analytical focus of this paper.

This undervaluation of competencies is not new and is often reflected in the working conditions imposed on workers. Historically, content moderation has relied on unpaid volunteer labour, establishing a pattern of uncompensated digital work that persists today. This precedent was set in the 1990s when tens of thousands of AOL volunteer community leaders managed chatrooms without payment, eventually leading to protests and a class action lawsuit that was settled in 2011 for 15 million USD \cite{matias2016going}. This historical reliance on unpaid labour has created ongoing tensions between platforms and moderators, as evidenced by collective action such as the Reddit moderator blackout, where volunteers protested changes to platform policies that affected their ability to perform moderation work effectively \cite{matias2016going}. At the same time, improvements in working conditions appear to be possible within BPOs, provided that there is third-party intervention, particularly from clients who can demand changes \cite{muldoon2025poverty}. Recent scholarship has proposed methods to improve moderation work by highlighting moderators' skills and the benefits their recognition could bring to platforms and society \cite{grimmelmann2015virtues, wilson2020hate, vaccaro2021contestability, fleisig2024perspectivist}. However, the psychological effects of content moderation work, particularly when moderating violent and hateful content, remain largely uncompensated and unrecognised \cite{steiger2021psychological, muldoon2024typology, gebrekidan2024content}.

\section{Positionality and Participatory Approach}

This study is grounded in a reflexive and participatory research approach, shaped by the authors' engagement in the sociopolitical conditions of platform labour and data rights. The first three authors are affiliated with the same university and NGO (PersonalData.IO, where they hold, respectively, the positions of director, research fellow and data scientist), approaching this work through their respective roles in research and advocacy. This study forms part of \emph{Data4Mods}'s project co-developed with the African Content Moderators Union (henceforth ACMU). 

Authors affiliated with ACMU form its steering committee, each holding specific roles: Kauna Ibrahim Malgwi \cite{timeKaunaMalgwi}, wellness and mental health lead; Mophat Okinyi \cite{timeMophatOkinyi}, educational trainer coordinator; Richard Mwaura Mathenge \cite{timeRichardMathenge}, committee administrator; James Oyange \cite{timeJamesOyange}, secretary; and Sonia Kgomo \cite{kgomo2025}, treasurer. Their combined experience spans the two main BPOs studied: Kauna Ibrahim Malgwi and Sonia Kgomo worked as Facebook content moderators through Sama (2019--2023 and 2021--2023 respectively), Mophat Okinyi held dual roles at Sama as data labeller and quality analyst (2019--2022), whilst James Oyange worked as a TikTok content moderator through Majorel (2022--2023), and Richard Mwaura Mathenge was employed by Majorel for three months in 2022. All members have since left these companies, and claim their contract terminations reflect patterns of retaliation linked to their advocacy: James Oyange's dismissal coincided with his advocacy efforts, whilst Kauna Ibrahim Malgwi and Sonia Kgomo were simultaneously terminated when their project was transferred from Sama to Majorel shortly before their contracts expired. 
It should be noted that all ACMU members hold university degrees in fields as diverse as psychology, human resources, public relations, international relations and management.

Adopting a reflexive ethnographic stance \cite{attia2017ing}, the authors recognise the influence of their positionalities on the research process, advocating for respecting the rights of affected communities marked by uneven power relations and regulatory asymmetries. Our engagement aimed not only at knowledge production but at producing practical means for workers to collect new information contributing to the union's collective strategies for improved working conditions. Building on recent contributions that foreground workers' voices \cite{dataworkersinquiry}, our approach goes further by positioning workers as active co-producers of scientific knowledge and researchers as actors engaged in social impact.

\subsection{The Participatory Process}

This research results from a seven-month participatory project between ACMU and PersonalData.IO (August 2024--March 2025). The project was conceived and developed as a partnership, avoiding the traditional researcher-participant model, and received external funding distributed upon budget agreement between the two organisations.

The collaboration was structured around weekly online meetings lasting between one and two hours, organised in two successive phases corresponding to the two data collection methods. In the first phase, dedicated to the map, the group established a common framework for data collection. Each member then collected information about different companies from their own knowledge and experience, which was subsequently harmonised and processed by the data scientist. The resulting map was discussed collectively during meetings to improve its readability and accuracy. The group then produced a press release and contacted journalists, with union members conducting interviews that led to published articles on the map's findings.

In the second phase, dedicated to the questionnaire, meetings shifted to a different mode of knowledge production. A shared repository was maintained where all participants contributed articles and references related to data work, which were discussed collectively to identify recurring themes in the literature and in workers' experiences. Union members recounted their experiences in the industry, with academics noting recurrent problems. Meeting minutes were taken and expanded through individual discussions, then analysed qualitatively through emerging themes, which formed the basis for questionnaire items. The group collectively discussed answer categories in light of workers' local context, for instance adjusting wage brackets to reflect local salary scales. Every union member reviewed the questionnaire, suggested improvements, and measured completion time. The group jointly produced the data management plan and consent form to ensure privacy protection. The union was then responsible for disseminating the questionnaire link through their network. Academics conducted the quantitative analysis and presented results to union members for collective interpretation, feeding directly into this article.

\subsection{Data Collection Methods}

This research uses mixed methods — desk research with map visualisation and questionnaire — that connect the global industry structure with systematic problems faced by workers. The mapping documents the structural conditions that obscure workers' experiences: subcontracting chains that diffuse accountability, corporate name changes that hide continuity, and geographical dispersion that prevents collective organising. The questionnaire then systematically documents whether the working conditions faced by union members are isolated incidents or industry-wide patterns across this dispersed workforce. The participatory meetings described above functioned as a third source of evidence: workers' accounts collected during sessions informed both the map's scope and the questionnaire's design, and are referenced in relation to the findings where they provide supporting qualitative evidence.

\subsubsection{Desk Research and Map Visualisation}

Desk research was conducted through the systematic collection of publicly available information from corporate websites of companies where workers had previously been, or are still, employed. Workers' experiential knowledge was essential at this stage, as companies often operate under different names, change over time, and conceal the nature of their operations. The aggregated information was processed to create a map using kepler.gl, a web-based geospatial data analysis tool, facilitating the construction of a mapping framework delineating workflows between corporate headquarters, subcontracting locations, and client companies. This approach combines qualitative analysis and data visualisation in alignment with mixed methods used to understand emerging labour markets.

\subsubsection{Questionnaire}

A cross-sectional online questionnaire was used to collect data on working conditions as experienced by workers across the industry, ensuring anonymity for respondents. Given the precarious and potentially retaliatory context in which respondents work, all questions were made optional, which accounts for variation in response rates across questions. The questionnaire covered employment history, contract conditions, and experienced challenges. This study does not claim statistical representativeness; it is explicitly exploratory. To our knowledge, no official database or statistics of data workers in Africa exists. Data analysis focused on descriptive statistics — frequency distributions, percentages, and visualisations — to examine key variables such as company identification, employment status, contract duration and renewal, working hours, salary ranges, and reported workplace challenges. Additional qualitative analyses were conducted to explore relationships between these variables and the companies where workers declared having worked. Given the small number of respondents per company for certain questions, these results should be interpreted with caution.
\section{Findings}

%%%%%%%%%%%%%%%%%%%%%%%%%%%% BPO MAP %%%%%%%%%%%%%%%%%%%%%%%%%%%%%%%
\subsection{Mapping the AI Industry \label{sec:map}}

\begin{figure}[h!]
    \centering
  \includegraphics[width=0.7\textwidth]{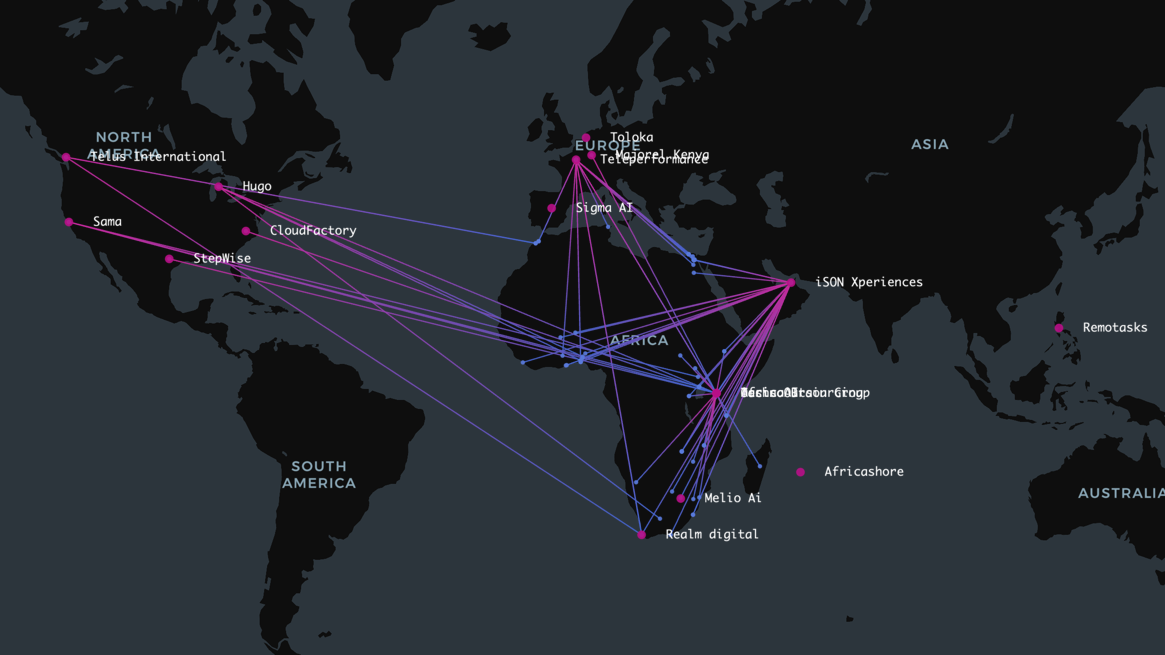}
  \caption{Map of the content moderation industry.}
  \label{fig:bpo-map}
\end{figure}

This section maps the BPOs forming the AI industry in Africa. Figure \ref{fig:bpo-map} presents the first visualisation of where data workers — encompassing both content moderators and data annotators or labellers — operate across the continent and the companies employing them. We identified 17 firms engaged in content moderation and data annotation, characterised by headquarter location, office locations in Africa and elsewhere, company size, and client base.

\subsubsection{Global Spread of the Industry and Location of Workforce}

Company headquarters are distributed globally across four main regions: North America (n=5, 29.4\%), Africa (n=6, 35.3\%), Europe (n=4, 23.5\%), the Middle East (n=1, 5.9\%) and South-east Asia (n=1, 5.9\%). This distribution reflects the global nature of the content moderation industry (with a notable concentration of headquarters within African markets themselves), alongside the emergence of regional African providers.

Operational presence differs from headquarters distribution. Ten companies (58.8\%) currently maintain active operations in Africa, operating across 43 of Africa's 55 countries (78.2\% of the continent). Among these, half have headquarters in Africa (Kenya, Republic of Mauritius, and South Africa) and half outside (United States, United Arab Emirates, France, and Spain). Concurrently, 11 companies (64.7\%) operate workforce centers in regions outside Africa, primarily concentrated in the Americas, indicating strategic global labour distribution. 

Some firms maintain exclusively African operations (e.g. Africashore), while others combine African centres with global operations (e.g. Teleperformance Nigeria). Others rely entirely on remote labour (e.g. Sigma AI, Toloka AI), outsourcing data labelling to African workers without local offices.

Notably, ACMU clarified that data work is often embedded within broader BPO operations, which sometimes describe their activities solely as ``call centre services'', obscuring their actual practices. Workers in these locations may be contractually designated as ``customer service representatives'' whilst performing data annotation, content moderation, and customer support simultaneously.

\subsubsection{Company Size Distribution}

Analysis of company size shows significant heterogeneity. Among firms with available size data (n=10), the largest operation is Africashore (3,047,956 employees), followed by Teleperformance Nigeria (410,000), Majorel Kenya (82,000), and Sigma AI (25,000). Smaller scale operations include Hugo (3,800), StepWise (1,001-5,000), and others with fewer than 200 employees (Realm Digital, Oasis Outsourcing, Melio Ai, AfricaAI). Overall, the industry is dominated by large multinational corporations alongside a smaller segment of specialised medium and small-scale providers; though workforce numbers remain unclear for companies relying on remote labour.

\subsubsection{AI Sectoral Demand for Data Work Services}

Available data (n=7 firms) show that content moderation and data labelling services support major AI platforms and diverse industries. Clients include technology companies such as Google (Hugo), cloud computing (Sigma AI), social media platforms (Hugo, Sigma AI), as well as automotive manufacturers (Africashore), banks, telecommunications firms, and retailers (Realm). 

Client base location reveals a South to North service flow: content moderation providers predominantly serve clients from developed economies (US, Europe) with some local/regional service provision, reinforcing Africa’s role as an offshore labour hub rather than a primary service market.

%%%%%%%%%%%%%%%%%%%%%%%%%%%% QUESTIONNAIRE: WORKING CONDITIONS %%%%%%%%%%%%%%%%%%%%%%%%%%%%%%%

\subsection{Working Conditions \label{sec:questionnaire}}

This section presents data workers' perspective of their working conditions across the mapped companies. Of 81 survey respondents, 67 (83\%) were unemployed at the time of data collection, while the remaining 14 (17\%) were still employed, which may indicate greater willingness among former workers to discuss their experiences. Concerning their role, 78 (96\%) identified as content moderators, 18 of which (23\%) also performed data labelling, and 3 (3.7\%) worked exclusively as data labellers. Geographically, of the 66 respondents who provided their location\footnote{Response rates varied across questions as participation was optional. All percentages are calculated on the basis of responses received for each item.}, the sample was concentrated in Kenya (n=62, 93.9\%), with small numbers based in Uganda (n=2, 3\%) and Ethiopia (n=2, 3\%). Throughout this section, we refer to all 81 respondents collectively as \textit{data workers}.

\subsubsection{In Which Companies Data Workers Work-ed?}

\begin{figure}[h!]
  \centering
  \includegraphics[width=0.6\textwidth]{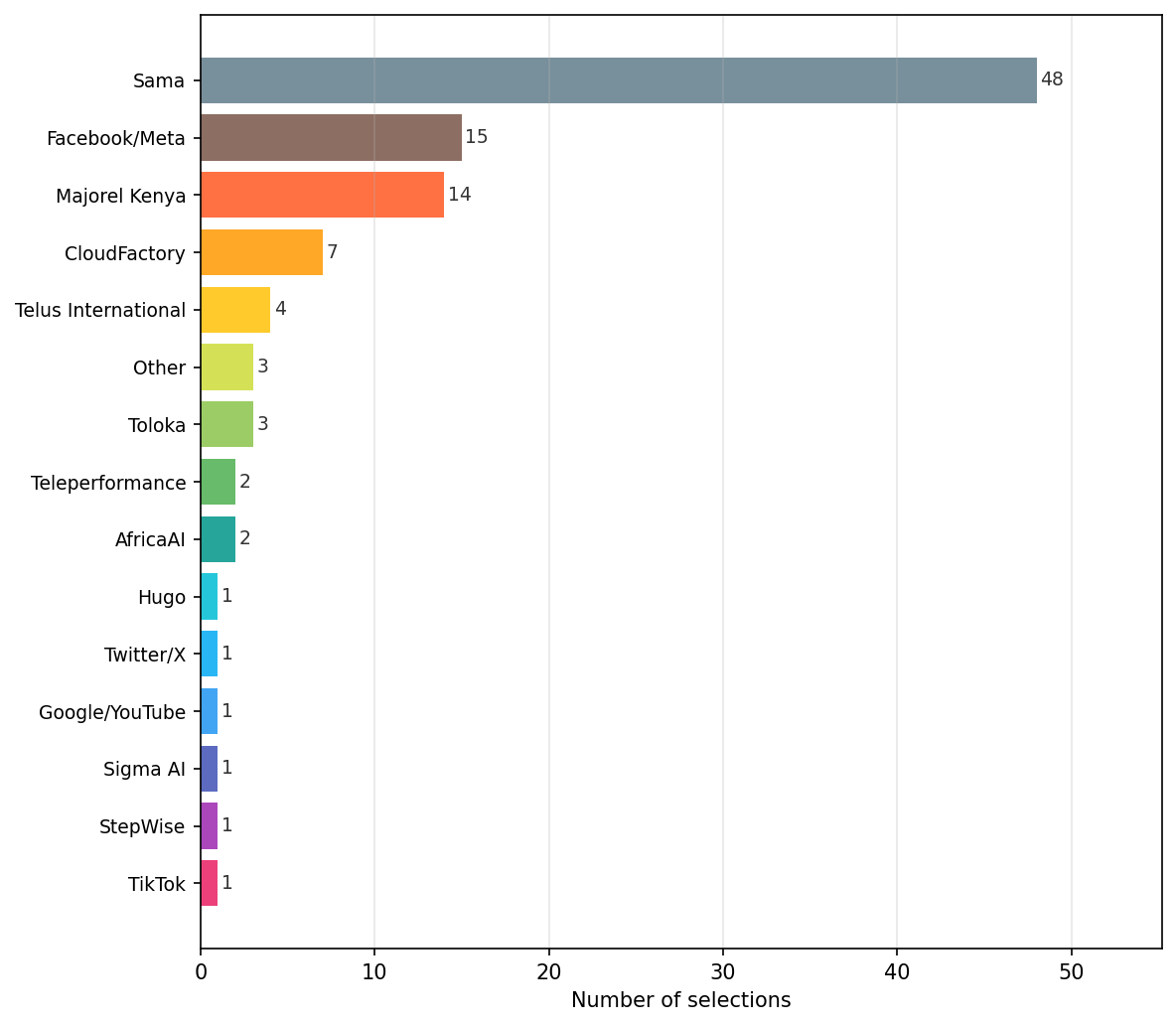}
  \caption{Company worked, as it appears in the contract (n=68 respondents; multiple selections possible).}
  \label{fig:companies}
\end{figure}

Of the 68 respondents who disclosed their current or former employer (Figure \ref{fig:companies}), the majority had worked for companies identified in the map (Section \ref{sec:map}). As workers could select multiple companies, total selections exceed the number of respondents. Multiple selection allowed them to express their full employment history. The three most represented employers were Sama (n=48), Facebook/Meta (n=15), and Majorel Kenya (n=14), followed by CloudFactory (n=7). The presence of social media platforms such as Facebook/Meta (n=15), TikTok (n=1), Google/YouTube (n=1), and Twitter/X (n=1) in the distribution requires careful interpretation. Some respondents selected both a BPO and a social media platform, reflecting that their work was performed for a platform company via a BPO subcontractor, which reflects the contested nature of the employment relationship between content moderators, BPOs, and platform companies \cite{njanja2023meta}, rather than necessarily indicating direct employment by those platforms. Four respondents selected Facebook/Meta only without any BPO, and their employment arrangements cannot be confirmed from the available data. Among the three respondents who selected ``Other'': Upwork and Remotasks were identified, pointing to the presence of additional platforms and crowdsourcing services not covered in our map.

In terms of work modality, of the 67 respondents who answered this question, most respondents (n=37, 55.2\%) worked exclusively on-site, while 27 (40.3\%) combined remote and in-person work; only 3 (4.5\%) worked fully remotely. This result shows that data work is predominantly location-dependent in this context of BPOs. On-site requirements facilitate the kind of intensive physical surveillance and spatial control more commonly associated with traditional factory workers than with new "digital" workers in the AI industry.

\subsubsection{Contractual Arrangements}

\begin{figure}[h!]
  \centering
  \begin{minipage}{0.48\textwidth}
    \centering
    \includegraphics[width=\textwidth]{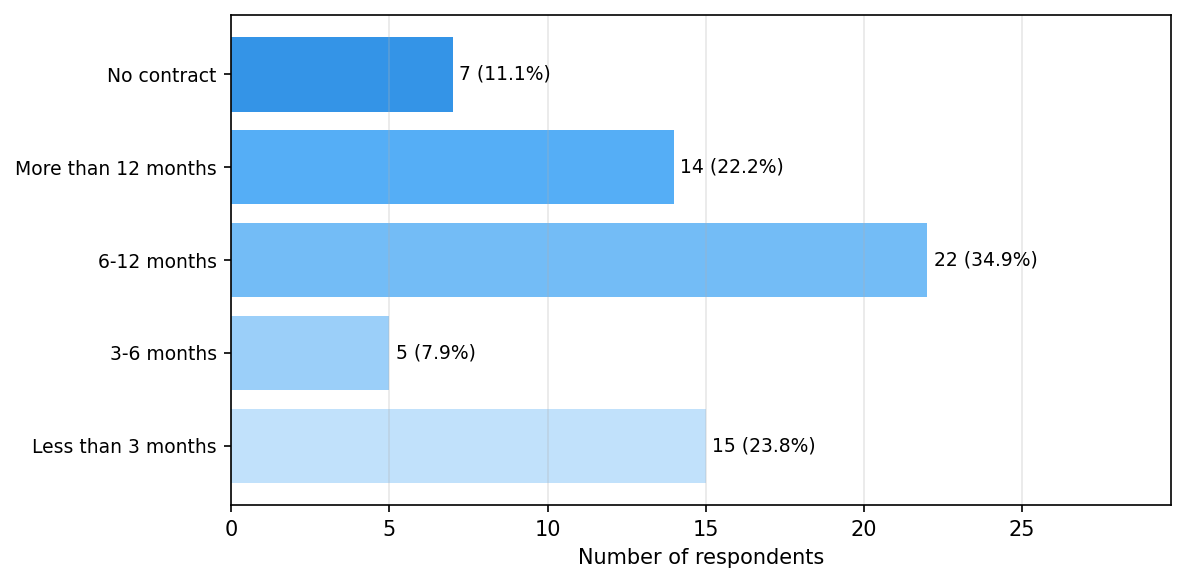}
    \caption{Contract duration (n=63).}
    \label{fig:contract-duration}
  \end{minipage}\hfill
  \begin{minipage}{0.48\textwidth}
    \centering
    \includegraphics[width=\textwidth]{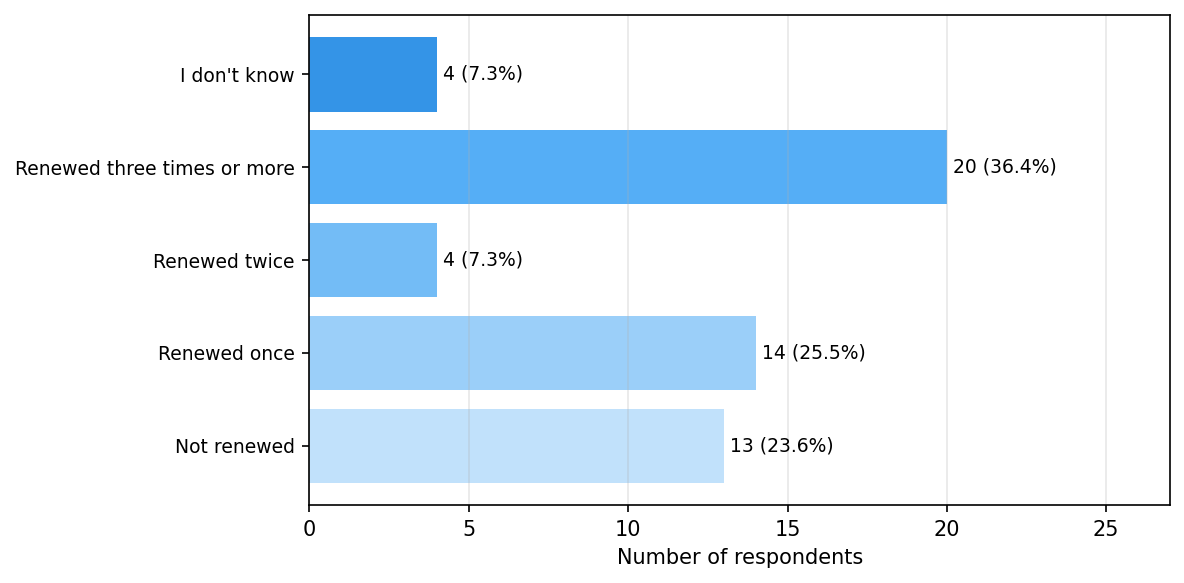}
    \caption{Contract renewal frequency (n=55).}
    \label{fig:contract-renewal}
  \end{minipage}
\end{figure}

Figure \ref{fig:contract-duration} shows contract duration among the 63 respondents who answered this question. The most common arrangement was 6--12 months (n=22, 34.9\%), followed by less than 3 months (n=15, 23.8\%) and more than 12 months (n=14, 22.2\%). Notably, 7 workers (11.1\%) reported having no contract at all. A smaller number held contracts of 3--6 months (n=5, 7.9\%). Figure \ref{fig:contract-renewal} shows renewal frequency among the 55 workers who answered this question. The largest group had contracts renewed three or more times (n=20, 36.4\%), whilst 14 (25.5\%) were renewed only once and 13 (23.6\%) received no renewal. A smaller number were renewed twice (n=4, 7.3\%) or did not know their renewal status (n=4, 7.3\%). Taken together, these patterns suggest that companies systematically rely on short-to-medium term contracts rather than offering permanent employment, with renewal practices varying considerably across workers, maintaining them in recurring uncertainty.

Looking at the companies represented among these respondents, Sama and Majorel Kenya appear as the primary drivers of these contractual patterns, though in distinct ways. Short contracts of less than 3 months were almost exclusively reported by Sama workers (n=14, 93.3\% of all less-than-3-month contracts), as were contracts renewed three or more times (n=18, 90.0\% of all such renewals). The pattern of non-renewal was also concentrated at Sama (n=10, 76.9\%) alongside Majorel Kenya (n=3, 23.1\%). Majorel Kenya workers, by contrast, were more concentrated in the 6--12 month category (n=5, 22.7\%)\footnote{Company-level percentages are calculated as the proportion of respondents affiliated with each company who reported a given response, out of all respondents from that company who answered the relevant question. As workers could select multiple companies, thins information should be interpreted as indicative rather than strictly representative of each company.}.

\subsubsection{Working Hours}

\begin{figure}[h!]
  \centering
  \begin{minipage}{0.48\textwidth}
    \centering
    \includegraphics[width=\textwidth]{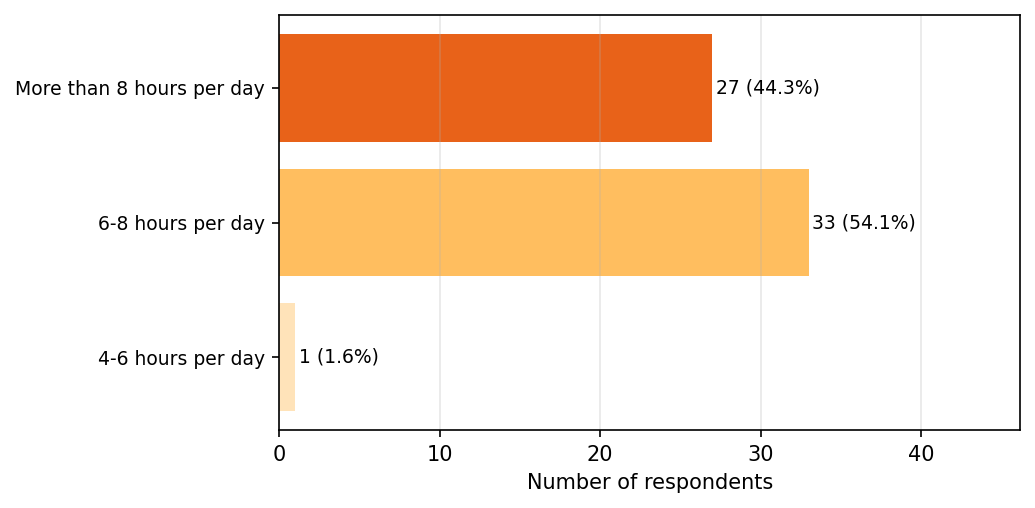}
    \caption{Contracted working hours per day (n=61).}
    \label{fig:contracted-hours}
  \end{minipage}\hfill
  \begin{minipage}{0.48\textwidth}
    \centering
    \includegraphics[width=\textwidth]{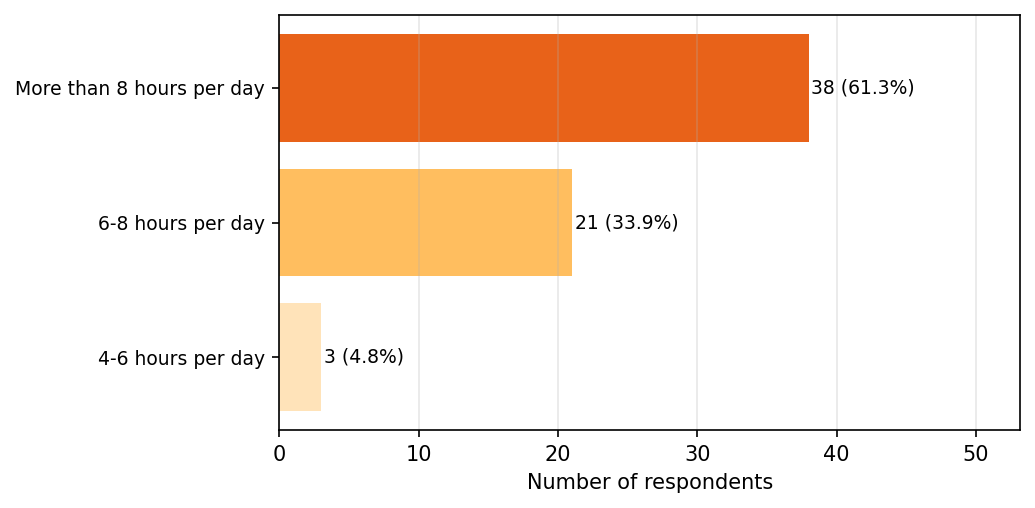}
    \caption{Actual working hours per day (n=62).}
    \label{fig:actual-hours}
  \end{minipage}
\end{figure}

Among the 61 respondents who answered the contracted hours question (Figure \ref{fig:contracted-hours}), most were contracted to work 6--8 hours per day (n=33, 54.1\%), whilst 27 (44.3\%) were contracted for more than 8 hours and only 1 (1.6\%) for 4--6 hours. Among the 62 respondents who answered the actual hours question (Figure \ref{fig:actual-hours}), 38 (61.3\%) reported actually working more than 8 hours per day, 21 (33.9\%) worked 6--8 hours, and 3 (4.8\%) worked 4--6 hours. This shift between contracted and actual hours suggests that overwork is routine in practice, exceeding the 8-hour standard set by national labour law in all surveyed countries \cite{kenyaWorkingHours, EthiopiaWorkingHours, UgandaWorkingHours}.

The gap between contracted and actual hours is most pronounced at Sama, where 11 workers contracted for 6--8 hours per day (45.8\% of Sama workers contracted for that range) reported actually working more than 8 hours. At Majorel Kenya, most workers were contracted for 6--8 hours (n=8, 80.0\%) and the majority actually worked within that range.

Our observations additionally show that there was an important difference in the problems faced by workers according to their position in the company. S, content moderator, reported arbitrary schedule changes, nine-hour working days with breaks limited to five minutes, making adequate meals or rest impossible under intense operational pressure. In contrast, M with a managerial position worked 9 hours per day with breaks totaling to an hour. He afforded better working breaks but required him to enforce productivity pressures on his team. He was responsible to enforce a quota of 50 tasks per day and maintain team productivity through direct pressure on subordinates.

\subsubsection{Salary}

\begin{figure}[h!]
  \centering
  \includegraphics[width=0.48\textwidth]{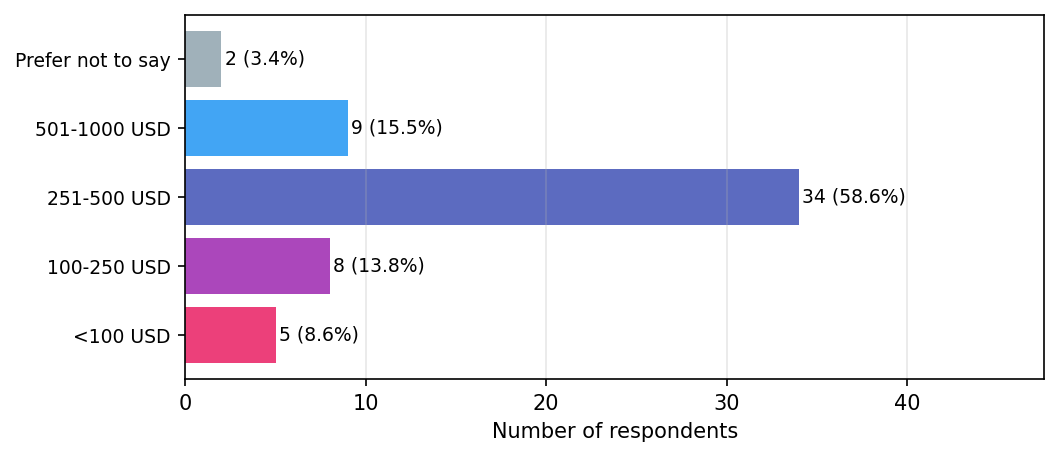}
  \caption{Monthly salary distribution (n=58).}
  \label{fig:salary-distribution}
\end{figure}

Among the 58 respondents who answered the salary question (Figure \ref{fig:salary-distribution}), the majority earned between 251--500 USD per month (n=34, 58.6\%), whilst 9 (15.5\%) earned between 501--1000 USD and 8 (13.8\%) earned between 100--250 USD. A smaller number reported earning less than 100 USD per month (n=5, 8.6\%), and 2 (3.4\%) preferred not to disclose their salary. Kenya's national average monthly earnings stand at 590 USD \cite{ceicData}, provided here as contextual reference only: the majority of respondents earned below this threshold, with 72.4\% (n=42) reporting monthly earnings of 500 USD or less.

Sama presents the most heterogeneous wage distribution, concentrating both the lowest earners and all 9 respondents earning above 500 USD per month. Majorel Kenya respondents were concentrated in the 251--500 USD range (n=7, 87.5\%), whilst the two CloudFactory respondents both reported earnings below 250 USD, though the very small sample warrants caution.

\subsubsection{Challenges}

\begin{figure}[h!]
  \centering
  \includegraphics[width=0.7\textwidth]{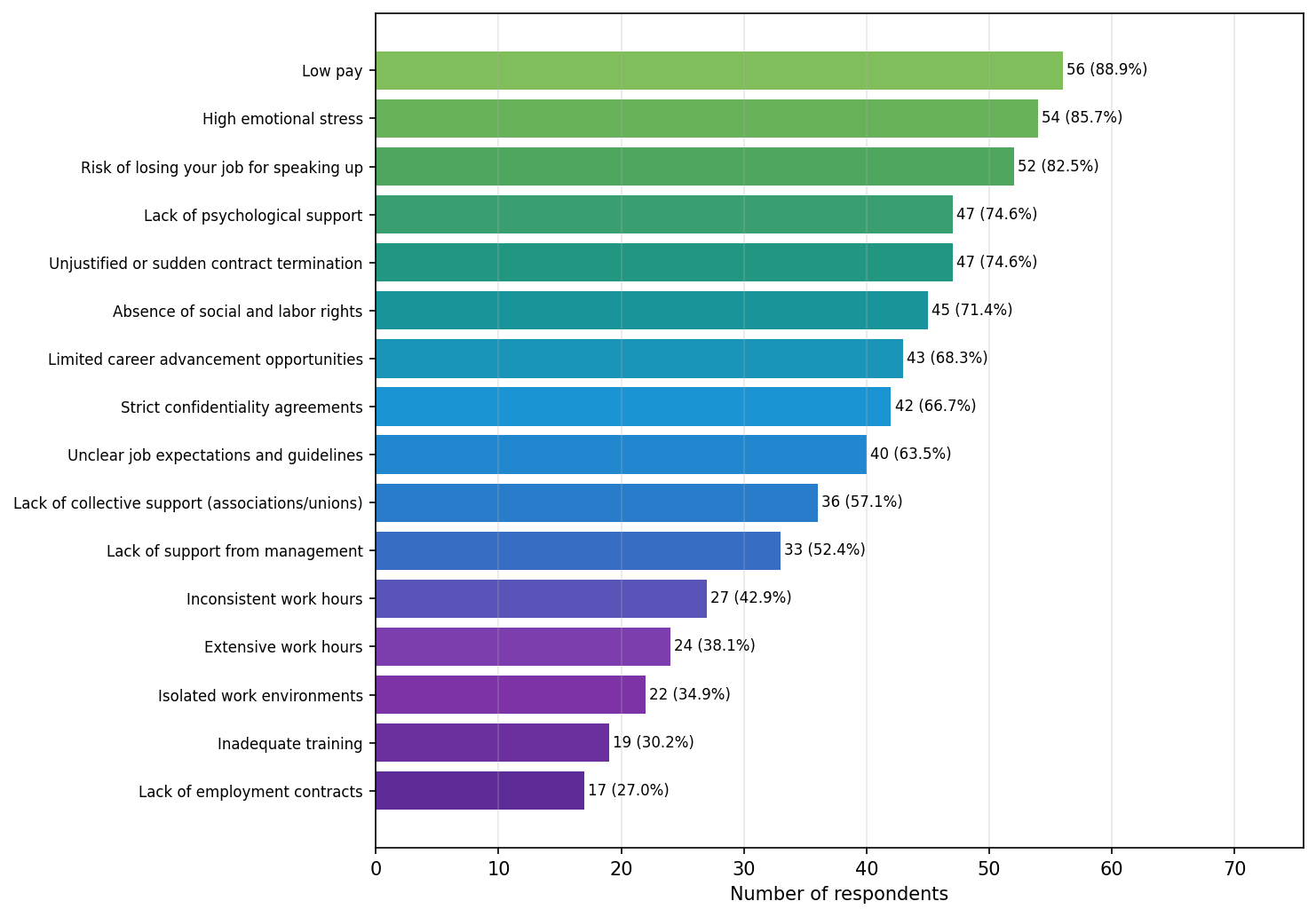}
  \caption{Reported challenges among all workers (n=63).}
  \label{fig:challenges}
\end{figure}

Among the 63 respondents who answered the challenges question (Figure \ref{fig:challenges}), low pay (n=56, 88.9\%) and high emotional stress (n=54, 85.7\%) were the most frequently reported challenges, followed by fear of reprisal for speaking up (n=52, 82.5\%), sudden contract termination and lack of psychological support (n=47, 74.6\% each), and absence of social and labour rights (n=45, 71.4\%). Limited career advancement opportunities (n=43, 68.3\%), strict confidentiality agreements (n=42, 66.7\%), and unclear job expectations (n=40, 63.5\%) were also commonly reported. {Taken together, these challenges point to two interconnected dimensions of precarity: a material and psychological one, where low pay and emotional stress are compounded by the absence of psychological support; and a structural one, where strict confidentiality agreements and fear of reprisal combine to silence workers contractually, whilst simultaneously fearing consequences for raising concerns, directly undermining their capacity for the collective voice that the struggle for recognition requires.}

Low pay and emotional stress are reported at high rates across both Sama (85.4\% and 83.3\% respectively) and Majorel Kenya (72.7\% each), suggesting these are sector-wide rather than company-specific issues. The structural silencing pattern — combining fear of reprisal and strict confidentiality agreements — is similarly present across both companies (Sama 79.2\% and 62.5\%; Majorel Kenya 72.7\% and 54.5\%).
\section{Discussion}

\subsection{Making the Invisible Visible}

\subsubsection{Supply Chain Structure and Working Conditions in Data Work BPOs}
Our research produces the first visual materialisation of the AI data work supply chain connecting the Global North and the Global South, making visible a chain of labour relations that has remained opaque due to the secrecy surrounding the sector.

Our analysis of 17 companies shows a globally distributed industry with extensive African operations: while headquarters are spread across regions, 35.3\% are based in Africa itself, suggesting the emergence of local companies alongside international firms establishing regional operations. These companies collectively operate across 43 of Africa's 55 countries (78.2\% of the continent), demonstrating the widespread geographic reach of data work across Africa. We show that data work flows from Global South to Global North, with African-based operations primarily serving as offshore outsourcing centers for North American and European companies. These outsourcing firms then provide content moderation and data annotation services for major AI applications developed by western technology companies, rather than serving local African markets. Our results confirm, with new evidence, segregative patterns of work provision in the Global South for Global North's AI development and consumption \cite{gebru2023digitalapartheid, elmahdielmhamdi2021}.

Data work demand spans multiple economic sectors, demonstrating that these services extend well beyond social media content moderation, the context on which most existing research has focused, and are integral to AI applications across a wide range of industries. This cross-sectoral presence is consistent with the broader role of BPOs as long-established organisations providing services across multiple sectors, from their origins as call centres \cite{apte1995global, miller2018call} to their current position as infrastructure for AI development. Our analysis identifies clients spanning automotive (Lear Corporation via Africashore), financial services (Standard Bank via Realm Digital), telecommunications (Vodacom via Realm Digital), retail (Spar via Realm Digital), consulting (PwC via Africashore), and cloud computing infrastructure (AWS, NVIDIA via Sigma AI), alongside technology platforms (Google, Meta via Hugo and Sigma AI). This cross-sectoral evidence raises the question of whether data work skills are genuinely transferable across industries or whether employers treat them as interchangeable for strategic reasons of labour substitutability.

The map provides a macro-level analysis of the industry's structural architecture, whilst the questionnaire data operates at the micro level, capturing the labour conditions within this structure. The questionnaire confirms and extends previous research \cite{cant2023fairwork, abdelkadir2025role, fairwork2025} on systematic precarity: short-term contracts with uncertain renewal, wages below national averages, routine overtime beyond contracted hours, and significant psychological and economic stress with limited institutional support. Taken together, the two levels of analysis show that the geographic and sectoral expansion documented in the map is sustained by poor working conditions consistent across companies and countries. This consistency suggests that precarity is not incidental but structural: the same supply chain logic that drives the offshoring of data work to Africa also drives the suppression of labour costs and worker protections.

\subsubsection{From Traditional BPOs to AI Data Work BPOs}

Our research complements other studies that explains that the organisational model of BPOs specialising in AI data work is similar to BPOs in other industrial sectors that practice the fragmentation of production work on an international scale \cite{ahmad2022moderating, tubaro2025does, anwar2025value}. The same asymmetries and dependencies between BPOs and their clients are therefore present, with direct consequences for how workers are treated and the conditions under which they operate.

BPOs also benefit from political decisions that can favour their development. For example, Kenya has chosen to export hundreds of thousands of Kenyans to countries such as the United Arab Emirates, Qatar and Poland through recruitment agencies, which have been accused of mistreatment and even exploitation \cite{semaforkenya}. Another solution put forward as a way to tackle youth unemployment is to promote job creation related to IT-enabled services and to ``unlock opportunities in the BPO services sector'' \cite{kenyagov}, leading to a positive policy towards companies such as Sama and Teleperformance. These factors prompt us to place our results within a broader international context. Indeed, since the establishment of BPOs, countries in the Global South are currently experiencing what countries in the Global North experienced in the 1980s with the rise of the service sector and service relationships, i.e. a shift in labour conditions from an industrial model of work — characterised by horizontal and vertical division of labour and continuous, direct supervision by first-level management—, to forms of organisation that impose both greater autonomy and constraints. BPOs are pure expressions of the tertiary industry of the Global North, now spread to countries in the South where labour laws and protections are less stringent or even non-existent, and where the tasks delegated to workers are considered to require little or no qualifications in the tech industry. However, these companies may hire university graduates in other domains from Global South countries to perform such work, revealing that offshoring is primarily motivated by labour cost arbitrage rather than actual skill and competences requirements. This is a question that directly concerns African Content Moderators Union, whose founding members are themselves educated and hold university degrees. This failure to take into account and fairly remunerate workers' level of education has led us to examine all the aspects of their daily work that can be considered as skills and competences required to perform their job.

\subsection{Content Moderation Work Extended Definition and Workers Competences}

\subsubsection{Broadening the Definition of Content Moderation in the Age of LLMs}

Organisational characteristics of BPOs (including the undervaluation of workers' expertise, the lack of recognition for their contributions to AI development, and the influence of colonial economic relationships) create particular working conditions that directly affect content moderators upskilling and recognition. Within BPOs, the exercise of expertise is not valued, and content moderators, although competent, are encouraged to remain superficial in the expression of their knowledge \cite{abdelkadir2025role}. Yet, content moderators, although they remain vague in terms of both the tasks they are required to perform and their career prospects (cf. challenges ``Limited career advancement opportunities'', ``Unclear job expectations and guidelines'' and ``Inadequate training'' identified by workers), find that their work is, in fact, undervalued and characterised by unpredictability. For example, content moderation workers often perform content processing tasks, even going so far as to carry out editorial work, ensuring that content complies with legislation and safeguards the platforms with regard to laws and advertisers \cite{roberts2014behind, gillespie2018custodians, gorwa2020algorithmic}. In addition, dominant models such as Llama (Meta), Gemini (Google), or ChatGPT (Open AI and Microsoft) belong to companies that own social media platforms that use the work of content moderators via BPOs \cite{siliconsavanna}. Social media data is used as training data for those LLMs \cite{MetaAI, MetaAI2, LinkedInAI, GoogleAI}. Thus, the work of content moderators, initially intended for social media, is being reused in the training of LLMs. In the era of massive development of LLMs, which are claimed to be an important step towards AGI~\cite{blili-hamelin2025position, El-Mhamdi_Hoang_Tighanimine_2025}, there is an urgent need to broaden the definition of content moderators in order to recognise the quantity and quality of their work, which will enable their true worth to be assessed. It is also a way of making more evident the phenomena of ``surplus labour'', i.e. labour produced by human data workers and that is not recognised for its true worth \cite{sambasivan2022all}) and ``precision labour'', i.e. ``the hidden and excessive labor involved in erasing the messy, ambiguous, and uncertain aspects of technology production'' \cite{zhang2025making}).

The expanded definition of data work allows us to propose, on the basis of our findings, an outline of professional competences that are not recognised as such. These are elements, tasks, behaviours and qualities that are naturalised, even though they require particular physical and cognitive commitment from workers. This naturalisation of know-how and social skills is comparable to what feminist theories have identified regarding the work performed by women, both in the private and professional spheres. Studies show that this phenomenon does not affect only women but all categories of workers who are denied the status of ``producers-reproducers'' \cite{kergoat2010rapport, monchatre2016salariat}. These workers occupy jobs where the required competence is at the intersection of a social position and a productive function. Data workers located in the Global South are part of these categories.

\subsubsection{Managing Unpredictability as a Competence}

Unpredictability mentioned below can also lead to adaptability, which can be regarded as a professional competence. The analysis of working hours demonstrates that BPOs consistently demand in practice extensive worker availability. 61.3\% of questionnaire respondents reported working more than 8 hours daily despite shorter contracted hours, demonstrating how temporal demands extend beyond formal agreements. Temporal availability represents one manifestation of flexible working arrangements, contributing to the unpredictability of work patterns, and ultimately to workers' precarious employment conditions. Nevertheless, this temporal availability can be considered an unrecognised competence. As studies focusing on service sector workers in direct contact with the public have demonstrated \cite{meda2005travail}, time availability signals commitment and motivation at work. It also demonstrates workers' ability to adapt to multiple tasks, take initiative, and exercise responsibility.

\subsubsection{Resilience as a Competence}

Another important element is that workers face as a challenge high emotional stress in combination with a lack of psychological support from their employers. This can be seen as another undervalued competence that both ACMU in this research project and content moderators in previous research \cite{abdelkadir2025role} advocate for: mental resilience. While commercial content moderation work frequently causes psychological problems amongst those who perform it, workers regard mental resilience as ``crucial expertise'' that should be recognised and prioritised by social media platforms and by extensions, BPO's \cite{abdelkadir2025role}.

\subsubsection{A Recognition of Competences for Fair and Equitable Pay}

There is a systematic devaluation of data workers' competencies leads to wages below national averages and precarious employment conditions characterised by short-term contracts and uncertain renewal prospects. This means that workers' skills and competences bear no relation to salary determination: despite the work demands, workers are underpaid and this is exacerbated by the fact that workers face unjustified or sudden contract termination.

In short, what the questionnaire responses show is that in practice, workers' responsibilities extend well beyond the poorly defined formal job descriptions that platforms typically provide to workers, which may result in deteriorating working conditions, encompassing both uncompensated labour and adverse psychological health outcomes. 

Thus, ACMU’s struggle takes on its full meaning and forms part of a social movement driven by professional motivations. ACMU is embracing the challenge identified by the questionnaire respondents, i.e. ``Risk of losing your job for speaking out'' (which ranks among the top three challenges). But their willingness to speak out, take risks and improve their own professional situation and that of their peers is precisely hampered by very real political and social obstacles.

\subsection{Actionable Recognition: Limitations and Perspectives}

Following the project, the union has drawn on this evidence to share knowledge among its members, raise public awareness through media outreach \cite{restofworld}, and produce a press release with journalist.

PersonalData.IO is also expanding the map with data gathered by researchers in France and the United States, towards a global open-access visualisation in their \href{https://personaldata.io/en/data4mods-2/}{website}. However, the union faces significant political and financial constraints. The union is facing problems with the local government that prevent them obtaining formal registration and blocking access to independent funding. Several spokespersons have been blacklisted from the industry, and at least three lack stable employment. One member has developed a mental health program for data workers \cite{mental},
whilst another now works for a different union. Despite these constraints, union members continue to advocate. Yet formal union registration, legal standing, and financial independence remain out of reach, leaving the struggle for recognition incomplete. Looking ahead, the map and questionnaire data could inform policy submissions, regulatory complaints, and advocacy before AI governance bodies. Should the union obtain formal registration, this evidence would strengthen claims for the reclassification of data work as a skilled occupation, and the claim is applicable wherever comparable supply chains exist.

Concerning more broadly research limitations, the map reflects a predominantly East and West African perspective grounded in the union's knowledge, inviting future work to extend the methodology to other regions. The questionnaire sample is concentrated at Sama (70.6\%) and geographically in Kenya (93.9\%), limiting cross-company and cross-country comparability; future research should build more balanced samples across BPOs and countries represented in the map. Response rates varied across questions due to the optional design; designating core questions as mandatory whilst keeping sensitive ones optional would improve comparability whilst protecting workers. Finally, the majority of respondents were former employees; given the sector's high contract rotation this gap may be less consequential than in more stable industries, but future research should make efforts to recruit currently employed workers to track conditions over time.
\section{Conclusion}

This paper has mapped the AI data work supply chain in Africa, and documented working conditions through a questionnaire within a participatory project, exploring patterns of precarity across companies and countries. By positioning workers as co-producers of knowledge, we have made visible labour relations that industry secrecy has kept opaque, and produced evidence that strengthens workers' claims for recognition. This work is part of a broader research tradition that goes beyond documentation alone. Data work and the workers who perform it have prompted researchers to respond through publications, inquiries, and participatory projects \cite{gray2019ghost, ajunwa2021tech, williams2022exploited, sambasivan2021everyone, whittaker2023origin, disalvo2024workers, dataworkersinquiry}. Our work is part of this tradition, which from a sociological perspective reflects not merely engagement but professional duty \cite{durkheim2019division}: the responsibility of researchers to address the social consequences of the systems they study. AI systems fundamentally depend on human workers whose labour our research demonstrates requires greater recognition and improved conditions. This presents an opportunity, and an obligation, for sociologists of work and computer scientists to collaborate in understanding and improving these human-AI workflows.

\bibliographystyle{ACM-Reference-Format}
\bibliography{biblio}

\section{Endmatter}

\subsection{Acknowledgements}
Our sincere thanks to David Décamps, Paul-Olivier Dehaye from PersonalData.IO and Ben Wray from Brave New Europe for their sustained contributions to communication, dissemination, and support throughout the duration of this project. We also thank the reviewers for their helpful comments in revising this paper.

\subsection{Research Ethics}
This research ethical framework addressed both questionnaire and data protection application within its broader participatory study design. Research information and consent sheets were developed collaboratively by PersonalData.IO and African Content Moderators Union to deploy the questionnaire, ensuring participants were fully informed about the study purpose, their data rights, and data collection procedures. All participation was voluntary and uncompensated. Questionnaire data was collected using an online tool. Subsequently, data was transferred to the NGO's secured cloud server in Switzerland, accessible only to project members for analysis purposes.

\subsection{Generative AI usage statement}
We used generative AI (Claude) for correcting English grammar (as the authors are non-native English speakers), and for producing plots in Python. No original research text, conceptual frameworks or empirical analyses were generated by AI.


\end{document}